\documentstyle[preprint,aps]{revtex}
\begin{document}

\title{\bf Echo in Optical Lattices: Stimulated Revival of Breathing Oscillations.}  
\author{ A. Bulatov$^1$, A. Kuklov$^2$, B.E.
 Vugmeister$^{1}$ and H. Rabitz$^{1}$} 

\maketitle

{\it $^1$ Department of Chemistry, Princeton University,
 Princeton, NJ 08544

\ $^2$ Department of Applied Sciences, The College of Staten Island,
CUNY, Staten Island, 

~~NY 10314}

\maketitle

\date{\today}
\maketitle
\begin{abstract}
We analyze a  stimulated revival (echo) effect for the breathing modes of the atomic oscillations in  optical lattices. 
The effect arises from the dephasing due to the weak
 anharmonicity being partly reversed in time by means of additional parametric excitation of the optical lattice. 
The shape of the echo response is obtained by numerically simulating the equation of motion for the atoms with subsequent 
averaging over the thermal initial conditions. 
A qualitative analysis of the phenomenon shows that the suggested echo
mechanism combines the features of both spin and phonon echoes.

\end{abstract}
\vspace{3mm}

\section{Introduction.}
 The coherent manipulation of the atomic center of mass motion 
in optical lattices \cite{JESSEN} by means of the nonstationary off-resonant dipole potentials has provided new experimental capabilities to study  dynamical systems with time-dependent potentials \cite{RAIZEN}-\cite{PHILLIPS}.
In the off-resonant coherent regime, the parametric non-adiabatic excitations of the optical lattice give rise to oscillations of the atomic momentum and coordinate distribution dispersions (breathing modes) 
\cite{GORLITZ,RUDY,PHILLIPS} and may be used for the manipulation of the coordinate or momentum dispersions of the atomic distribution by means of squeezing in phase space \cite{RUDY}. 

The squeezed states of matter in various systems have been
extensively discussed in connection with the possibility of overcoming the standard
quantum limit for noise imposed by vacuum fluctuations 
(see e.g. \cite{AVER,KIMBLE}). In an optical lattice, the squeezing effect can be significant in the classical regime producing classical squeezed states \cite{PRITCHARD1} and resulting in the reduction of thermal fluctuations.

Recently, the breathing modes of atomic oscillations in the
optical lattices have been observed experimentally with the use of the Bragg
scattering techniques based on the fact that the cross section of the Bragg
diffraction scales with the Debye-Waller factor which is related to the
dispersion of the atomic coordinate distribution \cite{GORLITZ,PHILLIPS}.
The observed decay of the oscillations may be due to both the dissipative and the dephasing effects. 

In classical oscillating systems with anharmonicity
the frequencies have a continuous distribution due to the energy-dependent corrections. This leads to the dephasing of small oscillations \cite{AVER}.
 Together with possible irreversible dissipation, the dephasing determines an apparent damping of the oscillations. 
The dephasing, caused by the nonlinearity effects, can lead to partial revivals under certain circumstances. Examples of such revivals are 
the phonon (see in \cite{MASON}), spin and photon echo effects 
\cite{HAHN}, \cite{MEYSTRE}.

One has to distinguish between the classical stimulated revival (echo)
caused by external perturbations applied to the oscillating system, and the quantum revival related to the discrete structure of the energy levels of the system which does not require any external disturbances \cite{AVER,PIT}. 
As will be shown below, these two revivals can have different time scales and therefore can be separated. 

In this paper, we show that an analog of the phonon echo effect exists for the breathing modes of atomic oscillations in the optical lattices.
However, the proposed echo mechanism possesses new features compared with the conventional phonon echo mechanism. 
First, the proposed mechanism is related to the oscillations
of the dispersion of the atomic distribution as opposed to the average 
coordinate and momentum oscillations involved in the phonon echo model \cite{MASON}. 
In contrast with the conventional
model of phonon echo \cite{MASON}, the excitation of the system in our model is achieved by means of the parametric modulation of 
the frequency of the atomic oscillations as opposed to applying a time-dependent external force to the atoms \cite{MASON}. 
As shown below, the suggested echo mechanism incorporates the features of 
both spin (photon)\cite{HAHN,MEYSTRE} and the phonon echo mechanisms 
due to the specific nature of the parametric excitation of the oscillations.
This effect can be employed to study the revival and damping mechanisms of the breathing modes in optical lattices. 

\section{ The Model and numerical results.}
An atom subjected to a off-resonant laser field experiences an energy shift of the ground state proportional to the intensity of the field. 
If the field is formed by a standing wave with large detuning, the effective potential for the atoms in the ground state is given by \cite{ZOLLER} 

\begin{equation}
U(x) = V[1 - \cos(2qx)],  
\label{eq:opt.lattice}	
\end{equation}
\noindent
where $x$ denotes the atomic center of mass coordinate,
 $V$ is the amplitude of the dipole potential
 (proportional to the intensity of the laser field), and
 $q$ is the wave vector of the laser field.
  Normally, the initial state of
 the system  is formed by Doppler cooling, so that 
the trapped atoms are in thermal equilibrium 
 near the minima of the optical potential and form an optical lattice.
In this paper we will consider the case of temperatures $T$ much
larger than the energy of the atomic oscillations
at the bottom of the potential. Consequently, no quantum
effects will be taken into account. On the other hand, the 
temperature should be low enough to insure that the atoms do not 
escape from the potential well.
If the magnitude of the optical lattice potential is time-dependent, then 

\begin{equation}
V = V_0 [1+k(t)],  
\label{eq:V}	
\end{equation}
\noindent
where $k(t)$ describes the parametric excitation of the lattice ($V_0$=const).
 Therefore, the effective potential can be approximated
 as an oscillator with weak anharmonicity and time-dependent
 frequency. The harmonic frequency corresponding to the potential (\ref{eq:opt.lattice}) is given by

\begin{equation}
\omega_0= 2q\sqrt{{V_0\over m}},  
\label{eq:omega_0}	
\end{equation}
\noindent
where $m$ is the atomic mass. We assume that the parametric excitation of the system occurs in the form of two short pulses 

\begin{equation}
k(t) = s_1\delta(t-t_1)+s_2\delta(t-t_2),
\label{eq:k(t)}
\end{equation}
\noindent
where $s_{1,2}$ and $t_{1,2}$ characterize the intensities and 
the instants when the pulses are applied to the optical lattice.
In what follows, we assume that $t_1=0$ and $t_2=t_p$.
The equation of motion is given by 

\begin{equation}
\ddot{\tilde{x}} = -\frac{1}{2} \sin(2\tilde{x})(1 + 2\xi_1 \delta(\tau) + 2\xi_2 \delta(\tau-\tau_p)),  
\label{eq:eq._of_motion1}
\end{equation}
\noindent
where we introduced the dimensionless variables

\begin{equation}
\tilde{x}(\tau)= q x(\tau),  ~~~\tau=\omega_0 t ,
\label{eq:abc1}
\end{equation}
\noindent
with $\xi_1=\omega_0 s_1/2$, $\xi_2=\omega_0 s_2/2$, $\tau_1=\omega_0 t_1=0$ and $\tau_2=\tau_p=\omega_0 t_p$. 

The physics of the echo effect in the optical lattice 
 can be understood as follows. 
We assume that initially the atomic system is in thermal equilibrium 
characterized by the values of coordinate and momentum dispersions $m\omega^2_0 <x^2>_T = <p^2>_T/m = k_B T$.
The first pulse initiates the squeezing of the atomic phase space distribution creating the classical squeezed states \cite{PRITCHARD1}. 
In harmonic systems, such squeezing results in oscillations of $<x(t)^2>$ and $<p(t)^2>$ with the frequency $2\omega_0$. 
The non-linearity leads to dephasing and therefore to the apparent 
damping of these oscillations.
Since the dephasing is reversible in time, one expects to observe at least a partial revival of the oscillations after the application of the second pulse to the system. A qualitative physical picture of the phenomenon will be discussed in the next section. Here we concentrate on the 
numerical results.

In order to study the echo effect, we performed a numerical simulation using the equation of motion (\ref{eq:eq._of_motion1}) with subsequent averaging of $\tilde{x}^2(\tau)$ over the initial conditions $\tilde{x}(0)$, $\tilde{\dot{x}}(0)$ distributed in accordance with the initial thermal 
ensemble. The results for the normalized average dispersion
$<\tilde{x}^2(\tau)>/<\tilde{x}^2>_T$ are presented in Fig.1. 
One can see the expected damping of the squeezing oscillations after 
each pulse due to the dephasing effect. The characteristic damping rate is increasing with an increase of the temperature. The most remarkable feature of the time evolution of $<\tilde{x}^2(\tau)>$ is the pronounced echo effect at $\tau \approx 2 \tau_p$. 
The revival time is approximately $2\tau_p$ according to the general
 considerations involved in the two-pulse phonon and the spin echo models. 
Note however, that the revival of $<\tilde{x}^2(\tau)>$ is only 
partial (approximately $0.25$ of the initial amplitude). 

In Fig.2, we present the dependence of the echo amplitude on the
intensity of the second pulse. The echo amplitude is very small for
small values of $\xi_2$, then it increases sharply in the intermediate region of $\xi_2$ and  saturates in the nonlinear regime.

\section{Qualitative analysis. Comparison with the phonon and spin
 echo.}
In order to clarify the mechanism of the proposed echo effect  and compare with the other echo mechanisms, 
we consider the approximate analytical solution of Eq.(\ref{eq:eq._of_motion1}) valid for the low temperature 
(small anharmonicity). In this case, it is sufficient to 
take into account only the first (quartic) correction to the effective harmonic potential. 
 Equation (\ref{eq:eq._of_motion1}) reduces to 

\begin{equation}
\ddot{\tilde{x}} = - \tilde{x}(1 +  2\xi_1 \delta(\tau) + 2\xi_2 \delta(\tau-\tau_p)) + \frac{2}{3} ~\tilde{x}^3 , 
\label{eq:eq._of_motion}
\end{equation}
\noindent
where we neglected the modulation of the non-linear term assuming that the 
amplitudes of the pulses $\xi_1$ and $\xi_2$ are sufficiently small.
For $\tau \neq 0, \tau_p$, the Eq.(\ref{eq:eq._of_motion}) has an integral of motion corresponding to the energy of the atomic system $\tilde{E}=E/4V_0$ 
(i.e., there are 3 different values of energy $E_k$, $k=0,1,2$ for times $\tau<0$, $0<\tau<\tau_p$ and $\tau>\tau_p$, respectively).

For small anharmonicity and  amplitudes of the external pulses, the
solution of the Eq.(\ref{eq:eq._of_motion}) for
$\gamma(\tau)=\tilde{x}^2(\tau)$ up to the second order terms in $E_k$
can be obtained in the form  

\begin{equation}\begin{array}{l}
\gamma(\tau)= C_k + [A_k\exp (2i\omega(\tilde{E}_k) \tau) +
 B_k\exp(4i\omega(\tilde{E}_k) \tau) +c.c], \\ \\
C_k = \tilde{E}_k + \frac{3}{4}\tilde{E}_k^2,
\label{eq:envelopek}
\end{array}\end{equation}
\noindent
where $k=0,1,2$. 
From Eq.(7), it follows that 
$A_k = \tilde{E}_k(1+{2\over 3}E_k) \exp(2i\phi_k)/2$ and $B_k =-A_k^2/6$, where $\phi_k$ corresponds to the phase of the atomic oscillations in different time intervals. Note that $A_k$ are the complex constants characterizing both the amplitude and phase of the atomic oscillations and $\omega(\tilde{E})$ stands for the effective frequency. This frequency is  determined in the limit $\tilde{E} \to 0$ as
\begin{equation}
\omega(\tilde{E}) = 1 - {1\over 2}\tilde{E}.
\label{eq:omgeff}
\end{equation}
\noindent
The solution given by Eqs.(\ref{eq:envelopek}) and (\ref{eq:omgeff}) corresponds to the first term of the asymptotic expansion  for the anharmonic oscillator in the limit of weak anharmonicity and large evolution time \cite{BOGOL}.

Each pulse in Eq.(\ref{eq:eq._of_motion}) results in a jump of $\dot{\gamma}$ and $\tilde{E}$, while $\gamma$ remains continuous. 
We obtain from Eq.(\ref{eq:eq._of_motion}) 

\begin{equation}
\dot{\gamma}(\tau_s^+)-\dot{\gamma}(\tau_s^-)=-4\xi_k\gamma(\tau_s);~~~ 
\tilde{E}(\tau_s^+)-\tilde{E}(\tau_s^-)=2\xi^2_k\gamma(\tau_s)-\xi_k \dot{\gamma}(\tau_s^-),
\label{eq:jump}
\end{equation}
\noindent
where $\tau_s=0, \tau_p$ and $\tau_s^+$ and $\tau_s^-$ denote the moment of time just before and just after each pulse and $k=1,2$. After the application of the 
first pulse, we obtain from Eqs.(\ref{eq:envelopek}) and (\ref{eq:jump}) 
for  small $\xi_1$

\begin{equation}
A_1=A_0+i\xi_1(\tilde{E}_0 + 2A_0)
\label{eq:A_1}
\end{equation}
\noindent
and

\begin{equation}
\tilde{E}_1=\tilde{E}_0-2i\xi_1(A_0-A^*_0).
\label{eq:E_1}
\end{equation}
\noindent
Combining Eqs.(\ref{eq:A_1}), (\ref{eq:omgeff}), and averaging over the random initial phase and the thermally distributed 
initial energy in the limit of low temperatures ($\Theta \equiv T/4V_0<<1$), 
we obtain the time evolution of the coordinate dispersion in the form

\begin{equation}
<\gamma(\tau)> = \Theta + \Theta [i\xi_1 \frac{\exp (2i\tau)}
{(1+i\tau/\tau_c)^2} + c.c.], 
\quad \tau_p > \tau > 0,
\label{eq:1pulse}
\end{equation}
\noindent
with the dephasing rate of oscillations defined by

\begin{equation}
\tau_c^{-1} = \Theta.
\label{eq:deph.rate}
\end{equation}
\noindent
Eq.(\ref{eq:1pulse}) indicates that the amplitude of the oscillations of the coordinate dispersion exhibits an apparent damping. 
As discussed above, this damping is caused
by the anharmonicity which results in  the dependence of the effective 
oscillation frequency on energy, and the thermal energy distribution 
leading to an averaging out of the oscillations. 
Therefore, this damping is reversible in time.  
The dephasing rate given by Eq.(\ref{eq:deph.rate}) is in a 
good qualitative agreement with the numerical results discussed above. Fitting the data presented in Fig.1 for $0 < \tau < \tau_p$ 
by Eq.(\ref{eq:1pulse}) gives the value $\tau_c^{-1} \approx 1.1 \Theta$ 
for the dephasing rate in our numerical simulations.

Now we will analyze the effect of the second pulse.
Applying Eq.(\ref{eq:jump}) for $\tau_s=0$ and $\tau_s=\tau_p$ consecutively, one obtains the amplitude $A_2$ in Eq.(\ref{eq:envelopek}) after the second pulse ($ t>\tau_p $)in the limit of small $\xi_2$

\begin{equation}
\tilde{A}_2(\tau_p) \approx \tilde{A}_1(\tau_p)(1+2i\xi_2)+i\xi_2\tilde{E}_1 - \frac{1}{3}i\xi_2 \tilde{E}_1 \tilde{A}^*_1(\tau_p)
\label{eq:2ampl}
\end{equation}
\noindent
where $\tilde{A}_{1,2}(\tau)=A_{1,2}{\rm e}^{2i\omega(\tilde{E}_{1,2}) \tau}$  correspond to the envelope solution for $\gamma(\tau)$ given by Eq.(\ref{eq:envelopek}). The $\tilde{A}$-vectors rotate with the frequencies $2\omega(\tilde{E}_{1,2})$ in the complex plane.
The energy after the second pulse is given by

\begin{equation}
\tilde{E}_2=\tilde{E}_1-2i\xi_2(A_1{\rm e}^{2i\omega(\tilde{E}_1) \tau_p}-c.c),
\label{eq:E_2}
\end{equation}
\noindent
with $A_1$ and $\tilde{E}_1$ defined by Eqs.(\ref{eq:A_1}) and (\ref{eq:E_1}). 

In order to obtain a qualitative picture of the echo effect, we 
disregard the energy change due to both pulses upon the 
frequencies $\omega(\tilde{E}_{1,2})$. 
Eq.(\ref{eq:2ampl}) indicates that the
amplitude $\tilde{A}_2(\tau_p)$ after the second pulse is a linear combination of the amplitude before the pulse $\sim \tilde{A}_1(\tau_p)$ and the amplitude $\tilde{A}_{2e}(\tau_p)=-i\xi_2\tilde{E}_1 \tilde{A}^*_1(\tau_p)/3$. Note, that $\tilde{A}_{2e}(\tau_p)$ is proportional to the complex conjugate of $\tilde{A}_1(\tau_p)$. 

In Fig.3, we present the two vector amplitudes
$\tilde{A_1}$ and $\tilde{A_1'}$ rotating with the frequencies 
$2\omega_1 >2\omega_1'$ in a complex plane. 
If the two vectors have the same phase $\phi=0$ at $\tau=0$ then 
the  phase difference between $\tilde{A_1}$ and
$\tilde{A_1'}$ is accumulated over the time interval $\tau_p$. 
The phases of the complex conjugate vectors $\tilde{A}_1^*$ and $\tilde{A}_1^{*'}$ are inverted with respect to $\tilde{A}_1$ 
and $\tilde{A}_{1}'$. 
The vector $\tilde{A}_1^{\prime}$ was rotating faster than the vector
$\tilde{A}_{1}$ and therefore $\tilde{A}_1^{\prime *}$ is delayed with respect to $\tilde{A}_1^{*}$. 
If the vectors continue rotating with the same frequencies   $2\omega_1$ and $2\omega_1'$, the phase difference between the vectors $\tilde{A}_1$ and $\tilde{A}_{1}'$ increases contributing to the dephasing effect. 
In contrast, the phase difference between $\tilde{A}_1^*$ and $\tilde{A}_1^{*'}$ vanishes at the time $\tau=2\tau_p$ giving rise to the echo analogous to the well known spin and photon echo effect \cite{HAHN}, \cite{MEYSTRE}.
Taking into account Eqs.(\ref{eq:envelopek})-(\ref{eq:jump}), (\ref{eq:2ampl})  one arrives at the echo contribution after thermal
averaging in the lowest order with respect to $\xi_1$, $\xi_2$ and $\Theta$
 
\begin{equation}
<\gamma(\tau)>_e =
  \Theta^2 ~\xi_1 \xi_2  
\frac{\exp(2i(\tau-2\tau_p))}{(1+i(\tau-2\tau_p)/\tau_c)^3} +c.c..
\quad \tau > \tau_p,
\label{eq:echo}
\end{equation}
\noindent
One can see that the amplitude of the echo is maximal
at the time $\tau = 2\tau_p $.
Note that the echo amplitude given by Eq.(\ref{eq:echo}) is proportional to the product $\xi_1 \xi_2$ in contrast to the dependence
 $<\gamma>_e \sim \xi_1 \xi_2^2$ typical for the phonon echo mechanism 
(for the small $\xi_1$, $\xi_2$) as a response to the external resonant
forces \cite{MASON}. However, the spin echo type mechanism described
above also leads to the terms $ \sim \xi_1 \xi_2^2$ omitted from Eq.(\ref{eq:2ampl}).  

According to the conventional model, the phonon echo originates from the change of the rotation frequency of the $\tilde{A}$-vectors due to the external pulses. 
This mechanism is also present in our model if we take into account the change of the energy $\tilde{E}$ in each time interval. 
In this case, the main contribution comes from the second term on the 
r.h.s. of the Eq.(\ref{eq:2ampl}). 
However, this effect is proportional to 
$\xi_1 \xi_2^2$ and vanishes for small pulse intensities. Analytical calculations in this case become rather cumbersome and we do not present them here. 
Since we disregard the change of the rotating frequency due to the second pulse, the approximation leading to Eq.(\ref{eq:echo}) is valid if 
$\Delta \tilde{E}_2 \tau_p \approx \xi_2\tilde{E}_0\tau_p << 1$. 
This condition can only be satisfied in the limit 
$\xi_2 << 1$ for large evolution time ($\tau_p >> 1$). 
This is a consequence of the asymptotic nature of the proposed analytical 
solutions \cite{BOGOL}.

Note, that the echo effect analyzed above is a stimulated 
classical revival of breathing oscillations in contrast to the effect of
quantum revival \cite{PIT} caused by the discrete nature of the energy levels
in quantum systems. We estimated a characteristic time of the quantum
revival $t_r$ (or $\tau_r=\omega t_r$) by extending the method 
of Ref.\cite{PIT} to finite temperatures. The ratio $\tau_c/\tau_r$ is
$4 \hbar\omega_0/ 3 \pi T$ and therefore 
$\tau_r >> \tau_c$ in the classical limit. 
Since the classical echo mechanism described above takes place in the time domain $\tau \approx 2 \tau_p$, the quantum revival will not contribute to the effect if $\tau_p < \tau_r$. 
For the value of $\Theta=0.03$ used in our numerical simulations 
and assuming that the number of levels in the potential well 
(\ref{eq:opt.lattice}) is $N \approx 100$, we obtain 
$\tau_c \approx 30$ and $\tau_r \approx 300$. Since $2\tau_p \approx 200$, 
the above mentioned condition is satisfied.

\section{Conclusion}
We investigated the dephasing and stimulated revival (echo) of the breathing oscillations in optical lattices subject to two-pulse
parametric excitation created by the abrupt change of off-resonant dipole optical potential induced by the laser field. 
The dephasing originates from the energy dependent corrections to the oscillation frequency and due to the thermal distribution of energy. 
The shape of the echo response was obtained by numerical simulation using
the equation of motion for the atoms with subsequent averaging over the initial thermal distribution. 
An approximate analytical analysis demonstrates that the 
suggested echo mechanism incorporates the essential physics features 
of spin(photon) echo where echo effect is caused by the 
change of the phase of the oscillations due to the second pulse, as well as 
the phonon echo where the effect is caused by the change of the 
oscillation frequency. 
An optimal control analysis would be desirable (e.g. as proposed for
the spin systems\cite{MRI}) for optimizing the echo response 
in view of the rapid development of novel experimental techniques (e.g. \cite{PHILLIPS}, \cite{RAIZEN})for the coherent manipulation of the atomic motion in optical lattices.

\acknowledgments
The work at Priceton University was  supported by the National Science
Foundation. A.K. acknowledges support by the New York PSC-CUNY Research
Award Program.



\newpage

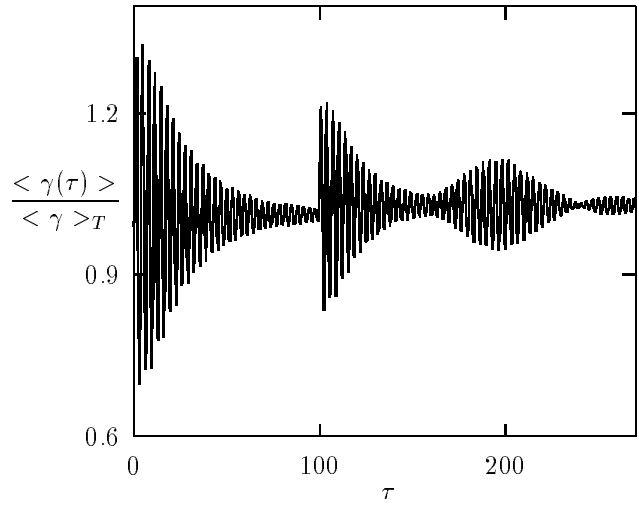
\begin{figure}
\caption{Fig.1. Numerical simulation of the echo effect discussed in
the text.
The parameters for the simulations are 
$\tau_p=105$, $\xi_1=0.2$, $\xi_2=0.1$, $\Theta=0.03$.}
\end{figure}

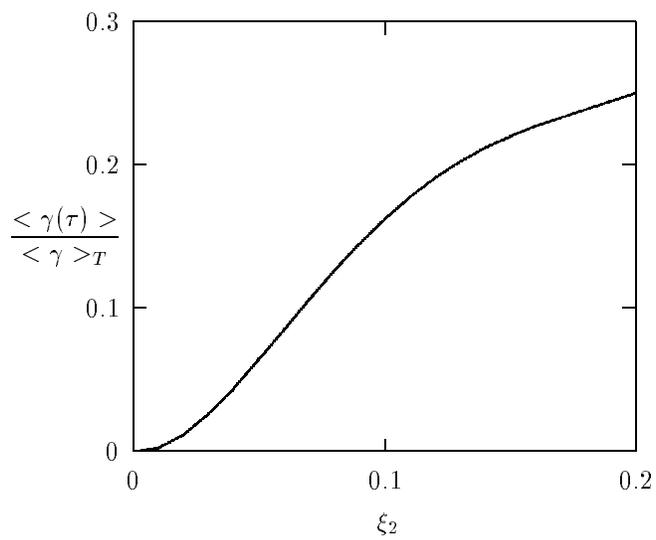
\begin{figure}
\caption{Fig.2. The maximum echo amplitude as a function of the
intensity $\xi_2$ of the second pulse.}
\end{figure}

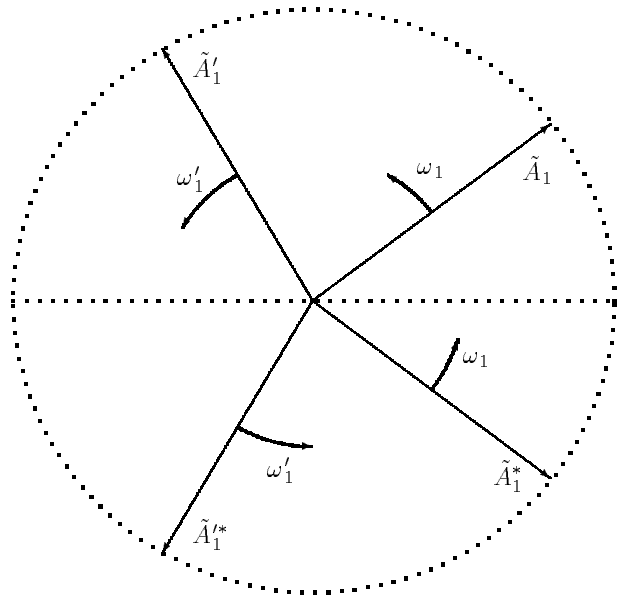
\begin{figure}
\caption{Fig.3. Illustration of the echo mechanism. Complex vectors 
$\tilde{A_1}$ and $\tilde{A_1^{\prime}}$  correspond to the different
values of the atom initial energy. For $0<\tau<\tau_p$, the vectors
rotate counterclockwise  with the  frequencies  $\omega_1^{\prime} > \omega_1$  and
accumulate the phase shift. At the moment $\tau=\tau_p$, the second
pulse is applied to the system  changing the oscillation
amplitude. The new amplitude $\tilde{A}_2$ is a supperposition of
$\tilde{A_1}$ and its
complex conjugate $\tilde{A_1}^*$
  For $\tau>\tau_p$, the phase difference between vectors
$\tilde{A_1}$ and $\tilde{A_1}^{\prime}$ decreases leading to the echo
effect.}
\end{figure}


\begin{thebibliography}{99}
\bibitem{JESSEN}
P.S. Jessen, C.Gerz, P.D. Lett, W.D. Phillips, S.L. Rolston, R.J.C. Spreew, and C.I.Westbrook, Phys.Rev.Lett {\bf 69}, 49 (1992);
G.Grynberg, B.Lounis, P.Verkerk, J.-Y. Courtois, and C.Salomon, 
Phys.Rev.Lett {\bf 70}, 2249 (1993);
A.Hemmerich and T.W. Hansch, 
Phys.Rev.Lett {\bf 70}, 410 (1993).
\bibitem{RAIZEN}
F.L.Moore, J.C.Robinson, C.F.Bharucha, B.Sundaram and M.G.Raizen, Phys.Rev.Lett {\bf 75}, 4598 (1995);
J.C.Robinson, C.F.Bharucha, K.W.Madison, F.L.Moore, B.Sundaram, 
S.R.Wilkinson and M.G.Raizen, Phys.Rev.Lett {\bf 76}, 3304
(1996); 
M.Raizen, C.Salomon and Q.Niu, Physics Today, 30 (July 1997).
\bibitem{GORLITZ}
A. Gorlitz, M. Weidemuller, T. W. Hansch, and A. Heimmerich, Phys.Rev.Lett. {\bf 78}, 2096 (1997).
\bibitem{RUDY}
P.Rudy, R.Ejnsman, and N.P.Bigelow, Phys.Rev.Lett. {\bf 78}, 4906 (1997).
\bibitem{PHILLIPS}
G.Raithel, G.Birkl, W.D.Phillips, and S.L.Rolston, Phys.Rev.Lett.{\bf 78}, 2928 (1997).
\bibitem{AVER}
I.Sh.Averbukh and N.F. Perel'man, Sov.Phys.Usp. {\bf 34}, 572 (1991);
I.Averbukh et al., Phys. Rev. {\bf A 50}, 5301 (1994).
\bibitem{KIMBLE}
H.J. Kimble and D.F. Walls, Eds., J.Opt.Soc.Am. {\bf B 4} (1987).
\bibitem{PRITCHARD1}
V.Natarayan, F.DiFilippo, and D.Pritchard, Phys.Rev.Lett. {\bf 74}, 2855 (1995).
\bibitem{MASON}
{\it Physical Acosutics}, Ed. by W.P.Mason, Volume XVI, Academic
Press, 1982.
\bibitem{HAHN}
E.L.Hahn, Phys.Rev. {\bf 80}, 580 (1950).
\bibitem{MEYSTRE}
P.Meystre and M.Sargent III, {\it Elements of Quantum Optics}, Springer-Verlag, 1990.
\bibitem{PIT}
L.P.Pitaevskii, Phys.Lett. A {\bf 229}, 406 (1997).
\bibitem{ZOLLER}
R.Graham, M.Schlautmann and P.Zoller, Phys. Rev A {\bf 45}, R19 (1992).
\bibitem{BOGOL}
N.N.Bogoliubov and Yu.A.Mitropolsky, {\it Asymptotic Methods in the
Theory of Nonlinear Oscillations} (Gordon and Breach,New York,1961)
\bibitem{MRI}
S.Conolly, D.Nishimura and A.Macovski, IEEE Trans. on Medical Imaging, 
{\bf MI-5}, 106 (1986).
\end{thebibliography}
\end{document}